\documentclass[fleqn,12pt,twoside]{article}
\usepackage{espcrc1}

\usepackage{graphicx,epsfig}
\usepackage[figuresright]{rotating}

\title{Exploring isospin, strangeness and charm distillation 
in nuclear collisions\thanks{Supported by DFG, BMBF, GSI and
BCPL}}

\author{M. Reiter\address[FFM]{Institut f\"{u}r Theoretische Physik, 
	Universit\"{a}t Frankfurt, 
	60054 Frankfurt, Germany}
	\address[msu]{Department of Physics, Michigan State University, East Lansing,
	MI, USA},
	E.~L.~Bratkovskaya\addressmark[FFM],
	M. Bleicher\addressmark[FFM], 
	W. Bauer\addressmark[msu],
	W. Cassing\address{Institut f\"{u}r Theoretische Physik, 
	Universit\"{a}t Giessen, 
	35392 Giessen, Germany},
	H. Weber\addressmark[FFM], 
 	and H.~St\"ocker\addressmark[FFM]\\}
	       
\begin{document}

\maketitle

\begin{abstract} {The isospin and strangeness dimensions of the Equation of
State are explored. RIA and the SIS200
accelerator at GSI will allow to explore these regions in compressed baryonic matter.
$^{132}$Sn+$^{132}$Sn and $^{100}$Sn+$^{100}$Sn collisions as well
as the excitation functions of $K/\pi$, $\Lambda/\pi$ and the centrality
dependence of charmonium suppression from the UrQMD and HSD transport models
are presented and compared to data. Unambiguous proof
for the creation of a 'novel phase of matter' from
strangeness and charm yields is not in sight.} \end{abstract}

\vspace{.5cm}

The properties of neutron-rich matter at high densities and temperatures are
largely unknown. Collisions of heavy rare isotopes provide an excellent tool to
explore this unknown region in the phase diagram of nuclear matter. The planned
Rare Isotope Accelerator (RIA) (up to $E_{\rm Lab}=500$~AMeV) and the SIS200 at
GSI (up to $E_{\rm Lab}=30$~AGeV) are the upcoming experimental facilities that
will access new energy density regions for these collisions allowing to
investigate the Equation of State (EOS) of compressed hadronic matter in
regions (2-10 $\rho_{\rm neutron}$) or even in directions ($n_{\rm
neutron}=n_{\Lambda}$) which have not been experimentally accessible before.

Among the most interesting open questions is the isospin and strangeness
dependency of the EOS, especially at high baryochemical potentials. This has
been the subject of intensive research, as it is not only important in the
study of radioactive nuclei but bears consequences on various important
astrophysical issues as well~\cite{Schaffner-Bielich:2000yj}. The EOS of
isospin asymmetric matter can be written as
$e(\rho,\delta)=e(\rho,0)+E_{sym}(\rho)\delta^2+O(\delta^4)$, where
$\delta\equiv(\rho_n-\rho_p)/(\rho_n+\rho_p)$ is the isospin asymmetry,
$e(\rho,0)$ is the energy density in isospin symmetric nuclear matter, and
$E_{sym}(\rho)$ is a parametrisation for the isospin dependence of the nuclear
symmetry energy~\cite{Wiringa:1988tp}. Parametrisations for $E_{sym}(\rho)$
fall into two main categories~\cite{Margueron:2001gx}: (I) $E_{sym}(\rho)$
rises with $\rho$, (II) $E_{sym}(\rho)$ falls with increasing density $\rho$.
The latter case has the interesting feature that symmetric matter may become
unstable at high densities.

The ($\pi^-/\pi^+$) ratio is particularly sensitive to the high density
behaviour of $E_{sym}$~\cite{Li}. This can be easily
understood qualitatively within the $\Delta$ resonance model for pion
production~\cite{Li}: Here first-chance independent nucleon-nucleon
collisions~\cite{Stock:xe} yield a primordial ($\pi^-/\pi^+$) ratio given by
$(5N^2+NZ)/(5Z^2+NZ)\approx(N/Z)^2$. Thus, the charged pion ratio yields
information on the isospin asymmetry (N/Z) in the dense participant matter,
which depends on the high density behaviour of $E_{sym}$~\cite{Li}. The
($\pi^-/\pi^+$) ratio can therefore be used to probe the EOS of neutron-rich
nuclear matter.

\begin{figure}[htb] \vspace{-.9cm}
\centerline{\psfig{figure=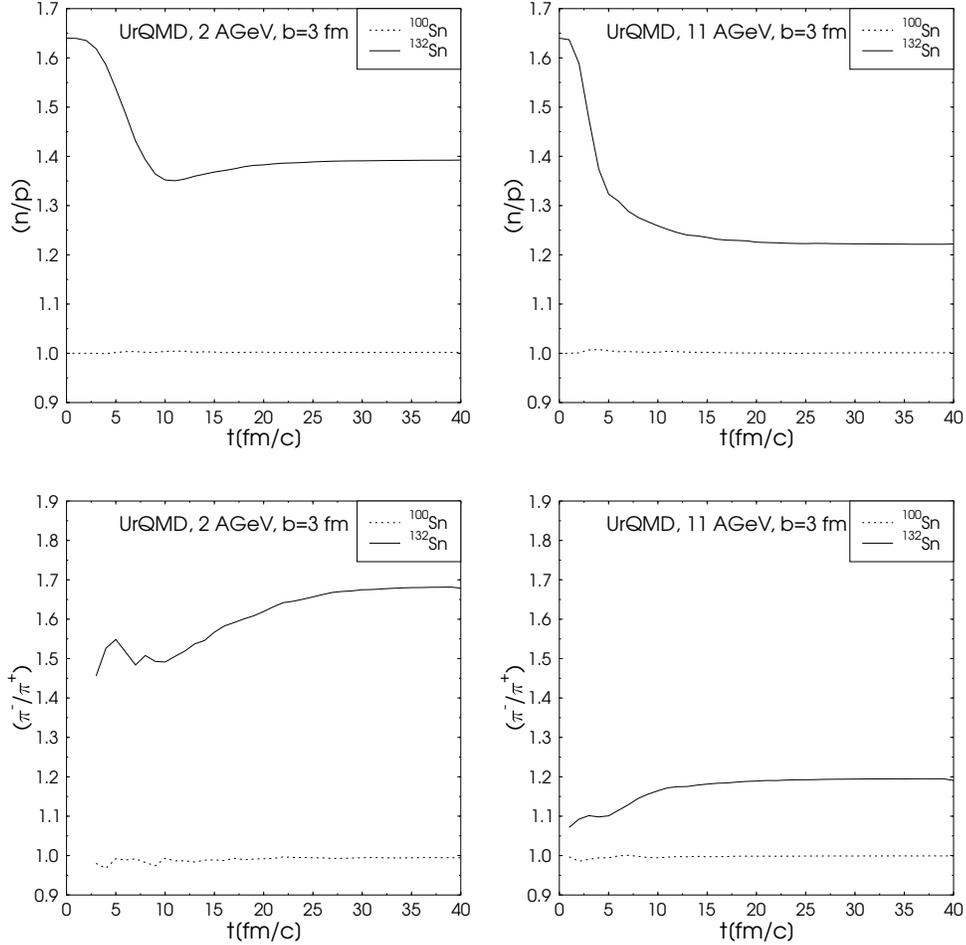,width=13.5cm}} 
\vspace{-1.5cm} 
\caption{Time evolution of the (n/p) (top) and $(\pi^-/\pi^+)$ (bottom) ratios
in central ($b=3$~fm) $^{132}$Sn+$^{132}$Sn and $^{100}$Sn+$^{100}$Sn
collisions at 2~AGeV (left panels) and 11~AGeV (right panels) as calculated
within the UrQMD model.}
\label{sn}
\vspace{-.5cm} 
\end{figure}

For our investigation of this effect the UrQMD model v1.2~~\cite{UrQMD1,UrQMD2}
is applied to $^{132}$Sn+$^{132}$Sn and $^{100}$Sn+$^{100}$Sn collisions at
$2$~AGeV and $11$~AGeV beam energy. Fig.~\ref{sn} shows the time evolution of
the (n/p) and ($\pi^-/\pi^+$) ratios. For $^{100}$Sn+$^{100}$Sn, both the (n/p)
and the $(\pi^-/\pi^+)$ ratios are roughly time independent and of order one,
as expected for the symmetric system. For the asymmetric system
$^{132}$Sn+$^{132}$Sn, the (n/p) ratio drops during the compression phase
from its initial value of $1.64$ and then saturates. At 2~AGeV, this
saturation occurs after 15~fm/c at a value of (n/p)~$\approx 1.4$, whereas at
11~AGeV the saturation occurs earlier (10~fm/c) and to a lower final
value of (n/p)~$\approx 1.2$.

The $(\pi^-/\pi^+)$ ratio rises slightly during the early phase of the reaction
and saturates rather later than (n/p) ($\sim 25-30$~fm/c at 2~AGeV and $\sim
15-20$~fm/c at 11~AGeV). The saturation values of $\sim 1.7$ (2~AGeV) and $\sim
1.2$ (11~AGeV) fall far below the first-chance $\Delta$ resonance model
prediction $(N/Z)^2\approx 2.7$. This is evident, as it is not the initial but
rather the local and dynamically changing (n/p) ratio that determines the
$(\pi^-/\pi^+)$ ratio. In addition, pion reabsorptions and rescatterings reduce
the sensitivity of $(\pi^-/\pi^+)$ to (n/p). At higher energies, the increasing
number of sequential nucleon-nucleon collisions drives the $(\pi^-/\pi^+)$
ratio towards $1$ via this mechanism.

For isospin-asymmetric matter, a distillation effect has recently been
proposed~\cite{DiToro:2003dq}, which is powered by chemical instabilities in
the liquid-gas phase transition driving the system towards isospin symmetry.
This effect is similar to the well-known {\it strangeness\/} distillation
mechanism in nuclear reactions, which has been proposed as a quark-gluon-plasma
(QGP) signal~\cite{Greiner:us}. The search for the QGP and its properties  has
been a driving force behind years of relativistic heavy ion physics.

The fact that the $\Lambda/\pi$ ratio and the $K^+/\pi^+$ ratio are found
experimentally to exhibit a maximum at $E_{\rm Lab}$=10-30 AGeV and drop to
half that value at 160 AGeV has raised speculations about the appearance of the
'new phase transition'. The NA49 Collaboration has recently started an energy
scan at the SPS. First results are available now at 40 and 80 AGeV
\cite{NA49_new,NA49_Lam,Misch02} and support these findings. Further studies
are underway at 30 and 20 AGeV \cite{SPS20}.

\begin{figure}[htb]
\vspace{-.1cm}
\centerline{\psfig{figure=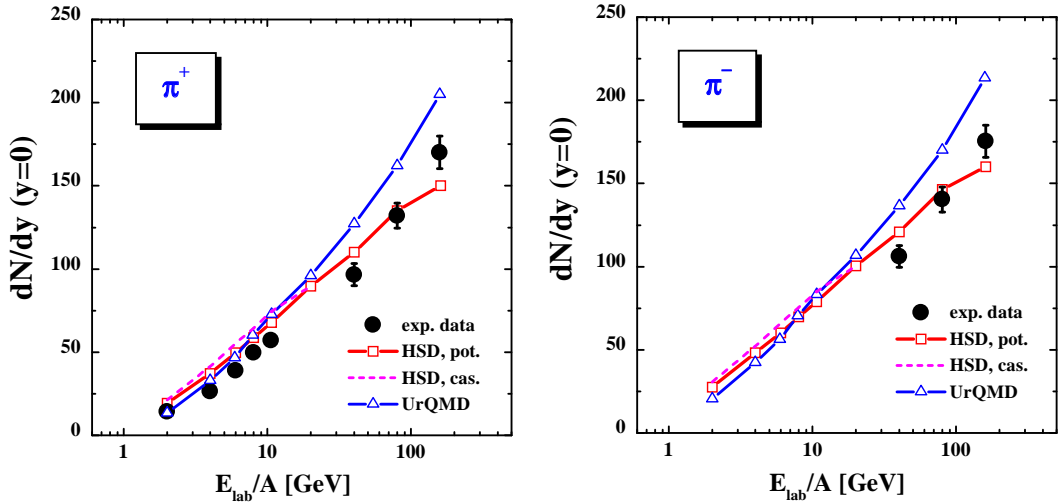,width=14cm}}
\vspace{-1cm}
\caption{Excitation function of $\pi^+$ and $\pi^-$ yields from central Au+Au
(AGS) or Pb+Pb (SPS) collisions  in comparison to experimental
data~\protect\cite{NA49_new,E866E917}. Solid lines with open triangles show
UrQMD results, solid lines with open squares and dashed lines stem from the HSD
approach with and without potentials, respectively~\protect\cite{prc}.}
\label{Fig4}
\vspace{-.5cm}
\end{figure}

Let us discuss these maxima in the excitation function on the basis of
'conventional' hadron transport theories -- do they exhibit analogous maxima?
In that case, the interpretation of the data as a QGP signal becomes
doubtful. The various hadron spectra are conventionally calculated with
non-equilibrium kinetic transport theory (cf.
\cite{CBRep98,Stoecker,Bertsch,Cass90,Koreview,WangSorge}). However, the
calculated kaon to pion ratio from central nucleus-nucleus collisions turns out
to vary by factors as large as 2 if different transport approaches are applied
\cite{WangSorge,HSD_exf,Jgeiss,WBS_K02}. Thus a unique interpretation of the
data is questionable so far.

Here we study the yields and ratios of protons, kaons, antikaons and hyperons
from Au+Au (or Pb+Pb) collisions in the energy range from 1~AGeV to
160~AGeV~\cite{prc}. The aim of this study is twofold: first, to find out the
systematic differences between two currently used transport approaches (UrQMD
\cite{UrQMD1,UrQMD2} and HSD \cite{CBRep98,Ehehalt}), and second, to compare to
related experimental data \cite{NA49_new,NA49_Lam,Misch02}. We provide
predictions for experimental studies in the near future \cite{SPS20}, which are
also of relevance for the new GSI-proposal \cite{GSIprop}.

The two different transport models employed, i.e.\ the UrQMD v1.3 (here an
updated version has been applied) and HSD approach have been used to describe
nucleus-nucleus collisions from SIS to SPS energies for several years. Though
different in the numerical realisation, both models are based on the same
concepts: strings, quarks, diquarks ($q, \bar{q}, qq, \bar{q}\bar{q}$)
accompanied by hadronic degrees of freedom. It is important to stress that both
approaches do not include any explicit phase transition to a quark-gluon plasma
(QGP). A common failure of both models in comparison to experimental data may
-- model independently -- indicate the appearance of a novel state of strongly
interacting matter.

\begin{figure}[htb]
\vspace{-.1cm}
\centerline{\psfig{figure=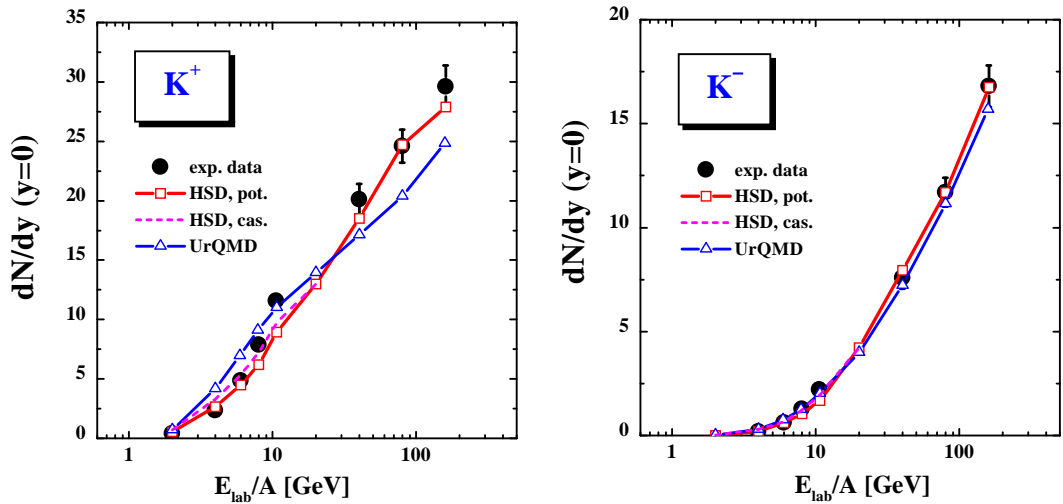,width=14cm}}
\vspace{-1cm}
\caption{Excitation function of $K^+$ and $K^-$ yields from central Au+Au (AGS)
or Pb+Pb (SPS) collisions in comparison to experimental
data~\protect\cite{NA49_new,E866E917}. Solid lines with open triangles show
UrQMD results, solid lines with open squares and dashed lines stem from the HSD
approach with and without potentials, respectively~\protect\cite{prc}.}
\label{Fig5}
\vspace{-.5cm}
\end{figure}

As a general overview on the $\pi^\pm$ abundancies in central Au+Au and
Pb+Pb collisions Fig.~\ref{Fig4} shows the $\pi^+$ and
$\pi^-$ multiplicities at mid-rapidity  as a function of the bombarding energy
in comparison to the available data~\cite{NA49_new,E866E917}.
Solid lines with open triangles show the results from the UrQMD
calculations while the solid lines with open squares and dashed lines stem from
the HSD approach with and without potentials, respectively. At lower AGS
energies the UrQMD model gives slightly less pions then HSD (with/without
potential), but  both models overpredict the mid-rapidity data. Around 10
AGeV is the "crossing point" of both transport calculations and at SPS
energies the tendency turns around: UrQMD gives more pions than HSD, so that
HSD is now in a better agreement with the experimental data.

The $K^+$ and $K^-$ multiplicities at mid-rapidity are depicted in
Fig.~\ref{Fig5} as a function of the beam energy in comparison to data from
Refs.~\cite{NA49_new,E866E917}: The $K^-$ abundancies are well described by
both transport models. The $K^+$ yield is slightly overestimated by UrQMD at
AGS energies and underestimated at SPS energies, whereas HSD is in reasonable
agreement with the data. Thus, Fig.~\ref{Fig5} demonstrates that an
underestimation of strangeness production is not the prevailing issue in
comparison to the recent data from NA49~\cite{NA49_new}. Both transport models
can roughly describe - within their systematic range of uncertainties - the
$K^\pm$ spectra and abundancies.

\begin{figure}[htb]
\vspace{-.1cm}
\centerline{\psfig{figure=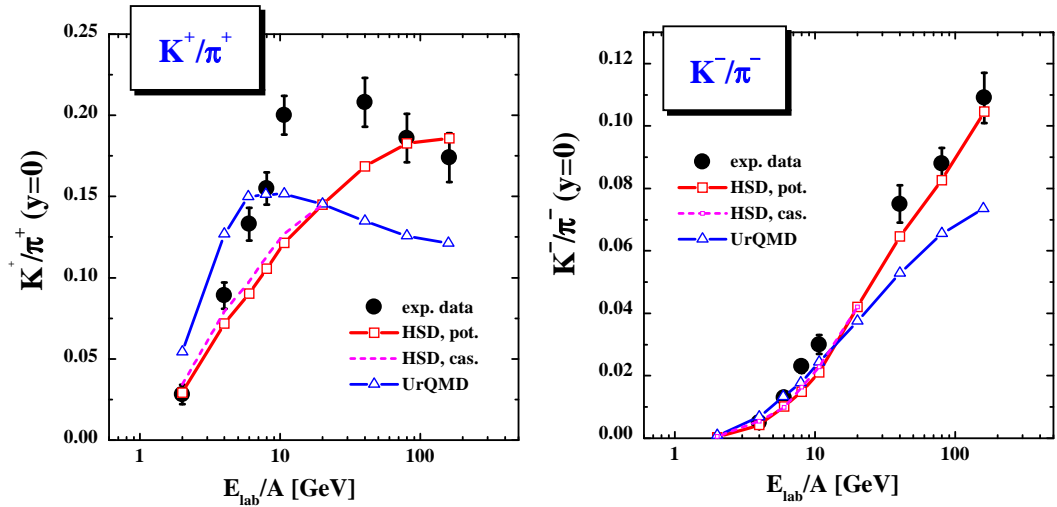,width=14cm}}
\vspace{-1cm}
\caption{Excitation function of the $K^+/\pi^+$ and $K^-/\pi^-$ ratios from
central Au+Au (AGS) or Pb+Pb (SPS) collisions  in comparison to experimental
data~\protect\cite{NA49_new,E866E917}. Solid lines with open triangles show
UrQMD results, solid lines with open squares and dashed lines stem from the HSD
approach with and without potentials, respectively~\protect\cite{prc}.}
\label{Fig6}
\vspace{-.5cm}
\end{figure}

Fig.~\ref{Fig6} contrasts the $K^+/\pi^+$ and $K^-/\pi^-$ ratios at mid-rapidity
as a function of energy for central collisions of Au+Au (AGS) or Pb+Pb (SPS)
with the data~\cite{NA49_new,E866E917}.  The excitation function of the
$K^-/\pi^-$ ratio is roughly reproduced by both transport models, the maximum
in the $K^+/\pi^+$ ratio seen experimentally is not described quantitatively by
either HSD or UrQMD. For the $K^+/\pi^+$ ratio both models give quite different
results. HSD gives a monotonous increase of this ratio with bombarding energy
(in qualitative disagreement with the data as pointed out in
Refs.~\cite{HSD_exf,Jgeiss}), whereas within UrQMD the ratio shows a maximum
around 10 AGeV and then saturates to a slightly lower value towards RHIC
energies, thus reproducing the behaviour of the data qualitatively. In view of
Figs.\ \ref{Fig4} and \ref{Fig5} this failure is not primarily due to a
mismatch of strangeness production, but more due to an insufficient description
of the pion abundancies.

This overestimation of pion abundancies can be attributed to a number of
reasons: One problem of the transport approaches used here is that detailed
balance is not implemented for $n\leftrightarrow m$ transitions with $n,m \ge
3$~\cite{Bravina}. Thus multi-particle collisions might change the dynamical
picture accordingly and lead to 'shorter' chemical equilibration times
\cite{Cass02_antip}. In fact, the importance of
$3\leftrightarrow 2$ transitions has been explored in an extended HSD
transport approach~\cite{Cass02_antip} for antiproton reproduction by
meson fusion for $A + A$ collisions at the AGS and SPS. In order to achieve a
more conclusive answer from transport studies multi-particle interactions
deserve further investigation in future generations of transport codes.

\begin{figure}[htb]
\vspace{-.1cm}
\centerline{\psfig{figure=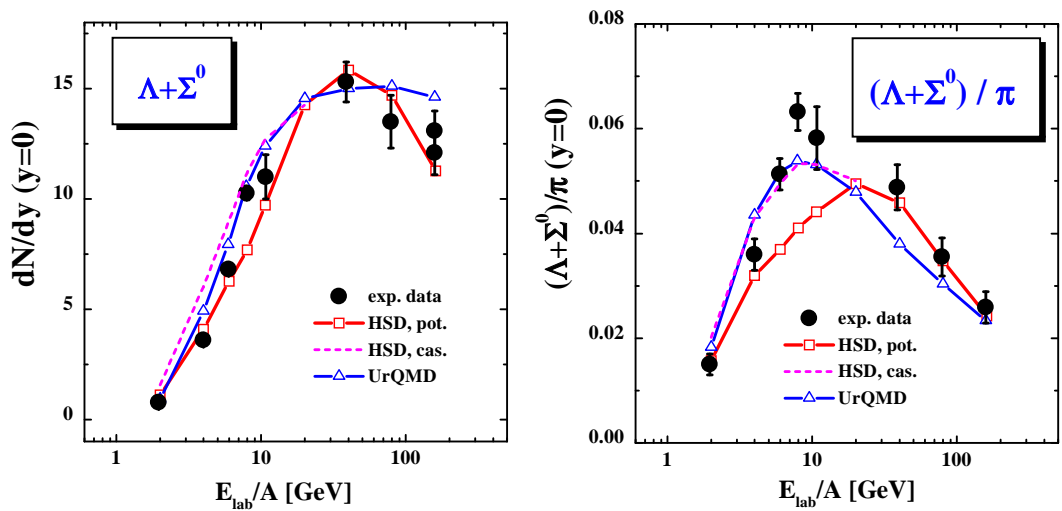,width=14cm}}
\vspace{-1cm}
\caption{Excitation function of the $\Lambda+\Sigma^0$ yields and
$(\Lambda+\Sigma^0)/\pi$ ratio from central Au+Au (AGS) or Pb+Pb (SPS)
collisions  in comparison to experimental
data~\protect\cite{NA49_new,E866E917}. Solid lines with open triangles show
UrQMD results, solid lines with open squares and dashed lines stem from the HSD
approach with and without potentials, respectively~\protect\cite{prc}.}
\label{Fig7}
\vspace{-.5cm}
\end{figure}

Another reason for the overestimation of the pion yield might be that the pions
in both transport models are treated as 'free' particles, i.e. with their
vacuum mass. Lattice QCD calculations and effective Lagrangian approaches like
the Nambu-Jona-Lasinio (NJL) model however, indicate  an increase in the pion
mass with temperature and density. So the overestimation of the pion yields in
HSD and UrQMD might be a signature for a dynamically larger pion mass.
Moreover, in-medium changes of the strange hadron properties, as known from
experimental studies at SIS energies, may also show up in the compressed
baryonic matter encountered in the AGS and SPS energy range. Thus, including
all medium effects simultaneously in a consistent way might provide a more
conclusive interpretation of the ratios in Figs.\ \ref{Fig6} and \ref{Fig7}. 
However, such calculations require a precise knowledge about the momentum and
density dependence of the hadron self-energies which is not available so far.

Fig.\ \ref{Fig7} shows the excitation functions of $\Lambda+\Sigma^0$-hyperons
(left) at mid-rapidity as a function of the bombarding energy for central
collisions of Au+Au (AGS) or Pb+Pb (SPS) in comparison to
data~\cite{NA49_Lam,Misch02,E891Lam,E896Lam,Antiori}.  Here both models compare
rather well with data. The $(\Lambda+\Sigma^0)/\pi$ ratios (right) at
mid-rapidity are underestimated slightly which should again be attributed to the
pion excess in the transport models (as discussed above).  Nevertheless, the
experimentally observed maxima  in the ratios are qualitatively reproduced by
both models, indicating that with increasing bombarding energy $s$-quarks are
more frequently produced within mesons ($\overline{K}, \overline{K}^*$) rather
than in associate production with baryons. A similar trend is also found in
statistical models~\cite{StatMod}.

What can we conclude from the common failures of the models studied in
comparison to related experimental data and does this failure provide evidence
for a new state of matter in view of a QGP? In view of the large 'systematic
uncertainties' in the discussed transport approaches we can quote an
insufficient accuracy in the description of the pion degrees of freedom by both
transport models. Thus, the deviations between theory and the data should not
be seen as unambiguous signals for any kind of new physics.

\begin{figure}[t!]
\vspace{-1cm}
\centerline{\psfig{figure=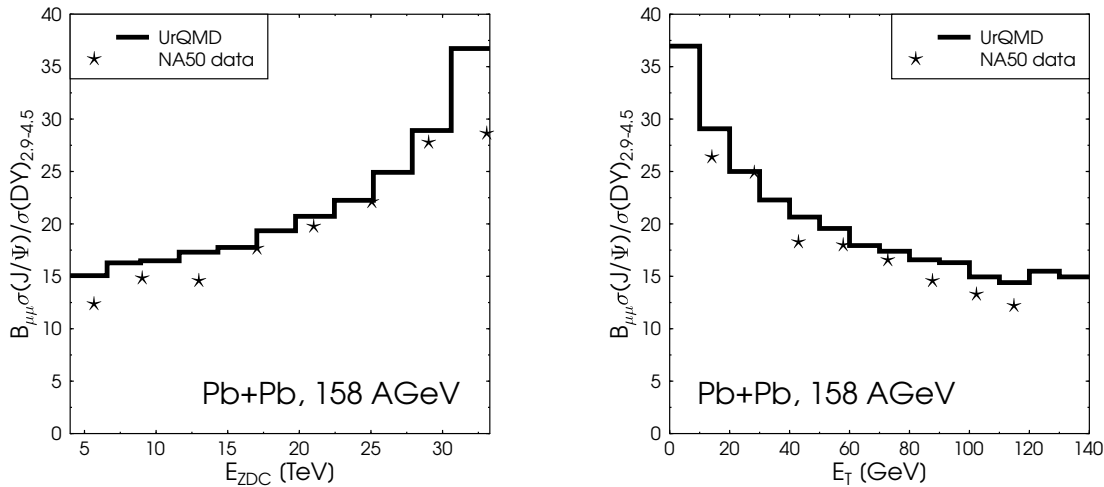,width=16cm}}
\vspace{-1.2cm}
\caption{The ratio of $J/\Psi$ to Drell-Yan production as a function of forward
energy $E_{\rm ZDC}$ (left) and transverse energy $E_{\rm T}$ (right) in
Pb+Pb collisions at 158 AGeV. Data points are preliminary results from the
year 2000 NA50 Pb+Pb run~\protect\cite{ramello}, the histograms show the 1999
UrQMD results~\protect\cite{Spieles:1999kp}.}
\label{jpsi}
\vspace{-.8cm}
\end{figure}

Another much-discussed possible QGP signal is charmonium production. Here, new
data on the $E_{\rm T}$ end $E_{\rm ZDC}$ dependence of $J/\Psi$ production in
Pb+Pb collisions at 158~AGeV from the year 2000 NA50 run has recently become
available~\cite{ramello}. This data (one analysis), along with UrQMD
results~\cite{Spieles:1999kp} (which are in line with HSD
results~\cite{Cassing:1997kw}) is shown in Fig.~\ref{jpsi}. Note that the
puzzling drop of the $J/\Psi$ to Drell-Yan production at high $E_T$ which was
present in the 'old', previously available NA50 data is no longer present in
the current NA50 data. The new NA50 data fall very close to the published UrQMD
calculations by Spieles et al.~\cite{Spieles:1999kp}. To our opinion, both the
$E_{\rm T}$ dependence and the $E_{\rm ZDC}$ dependence do no longer justify to
speak of an 'anomalous' $J/\Psi$ suppression in high energy heavy ion
collisions.

We have shown that the exploration of the isospin and strangeness dimensions of
the Equation of State opens fascinating new areas of research. Especially the
upcoming experimental possibilities like the planned Rare Isotope Accelerator
RIA and the SIS200 machine at GSI will allow to explore novel phenomena in
compressed baryonic matter in great detail. While conclusive unambiguous proof
for the creation of a 'novel phase of matter', e.\,g. Quark Gluon Plasma, from
strangeness and charm yields is still lacking, the hunt and quest for it has
provided new insights and even opened novel fields of research. It has become a
driving force behind modern high energy nuclear physics.

\section*{Acknowledgement}

We thank A.~Kostyuk, L.~Gerland, R.~Vogt, and L.~Csernai for many
fruitful discussions.


\begin{thebibliography}{9}
\bibitem{Schaffner-Bielich:2000yj}
J.~Schaffner-Bielich, M.~Hanauske, H.~St\"ocker and W.~Greiner,
astro-ph/0005490.
\bibitem{Wiringa:1988tp}
R.~B.~Wiringa, V.~Fiks and A.~Fabrocini,
Phys.\ Rev.\ C {\bf 38} (1988) 1010.
\bibitem{Margueron:2001gx}
J.~Margueron, J.~Navarro, N.~Van Giai and W.~Jiang,
nucl-th/0110026.
\bibitem{Li}
B.~A.~Li,
Phys.\ Rev.\ Lett.\  {\bf 88} (2002) 192701;
B.~A.~Li,
nucl-th/0301039.
B.~A.~Li,
Phys.\ Rev.\ C {\bf 67} (2003) 017601
B.~A.~Li,
Nucl.\ Phys.\ A {\bf 708} (2002) 365
\bibitem{Stock:xe}
R.~Stock,
Phys.\ Rept.\  {\bf 135} (1986) 259.
\bibitem{UrQMD1}
    S.A.~Bass et al.,
    Prog. Part. Nucl. Phys. {\bf 42}, 279 (1998).
\bibitem{UrQMD2}
    M.~Bleicher et al.,
    J. Phys. G {\bf 25}, 1859 (1999).
\bibitem{DiToro:2003dq}
M.~Di Toro et al.,
nucl-th/0301033.
\bibitem{Greiner:us}
C.~Greiner and H.~St\"ocker,
Phys.\ Rev.\ D {\bf 44} (1991) 3517.
\bibitem{NA49_new}
     S.V.~Afanasiev et al. (NA49 Collab.),
    Phys. Rev.~C. {\bf 66} (2002) 054902
\bibitem{NA49_Lam}
    A.~Mischke et al. (NA49 Collab.), J. Phys. G. {\bf 28}, 1761
    (2002).
\bibitem{Misch02} A.~Mischke et al. (NA49 Collab.),
    nucl-ex/0209002, to be published in Nucl. Phys.~A.
\bibitem{SPS20}
    The NA49 Collaboration, Addendum-10 to Proposal
    CERN/SPSC/P264.
\bibitem{CBRep98}
    W. Cassing and E. L. Bratkovskaya, Phys. Rep. {\bf 308}, 65 (1999).
\bibitem{Stoecker}
    H. St\"ocker and W. Greiner, Phys. Rep. {\bf 137}, 277 (1986).
\bibitem{Bertsch}
    G. F. Bertsch and S. Das Gupta, Phys. Rep. {\bf 160}, 189 (1988).
\bibitem{Cass90}
    W. Cassing, V. Metag, U. Mosel, and K. Niita,
       Phys. Rep. {\bf 188}, 363 (1990).
\bibitem{Koreview}
    C. M. Ko and G. Q. Li, J. Phys. G {\bf 22}, 1673 (1996).
\bibitem{WangSorge}
    F. Wang, H. Liu, H. Sorge, N. Xu, and J. Yang,
    Phys. Rev. C {\bf 61}, 064904 (2000).
\bibitem{HSD_exf}
     W.~Cassing, E.~L.~Bratkovskaya, S.~Juchem,
     Nucl.  Phys. A {\bf 674}, 249 (2000).
\bibitem{Jgeiss}
    J. Geiss, W. Cassing, and C. Greiner,
    Nucl. Phys. A {\bf 644}, 107 (1998).
\bibitem{WBS_K02}
     H. Weber, E.L. Bratkovskaya, and H. St\"ocker,
     Phys. Lett. B {\bf 545}, 285 (2002).
\bibitem{prc}
    H.~Weber, E.~L.~Bratkovskaya, W.~Cassing and H.~St\"ocker,
    Phys.~Rev.~C {\bf 67} (2003).
\bibitem{Ehehalt}
    W. Ehehalt and W. Cassing, Nucl. Phys. A {\bf 602}, 449 (1996).
\bibitem{GSIprop}
    http://www.gsi.de/GSI-Future/cdr/.
\bibitem{E866E917} 
    L. Ahle et al. (E866 and E917 Collab.),
    Phys. Lett. B {\bf 476}, 1 (2000);
    Phys. Lett. B {\bf 490}, 53 (2000).
\bibitem{Bravina}
    M. Belkacem et al.,
        Phys. Rev. C {\bf 58}, 1727 (1998);
    L.V. Bravina et al.,
        Phys. Lett. B {\bf 434}, 379 (1998);
    L.V. Bravina et al.,
        J. Phys. G {\bf 25},351 (1999);
    L.V. Bravina et al.,
        Phys. Rev. C {\bf 60}, 024904 (1999).
\bibitem{Cass02_antip}
    W. Cassing, Nucl. Phys. A {\bf 700}, 618 (2002).
    R. Rapp and E. V. Shuryak,
    Phys. Rev. Lett.  {\bf 86}, 2980 (2001).
    C. Greiner and S. Leupold,
    J. Phys. G {\bf 27}, L95 (2001).
\bibitem{E891Lam} 
      S. Ahmad et al. (E891 Collab.), Phys. Lett. B {\bf 382}, 35 (1996);
      C. Pinkenburg et al. (E866 Collab.),
    Nucl. Phys. A {\bf 698}, 495c (2002).
\bibitem{E896Lam} 
       S. Albergo et al., (E896 Collab.),
	Phys. Rev. Lett. {\bf 88}, 062301 (2002).
\bibitem{Antiori}
    F. Antinori at al. (WA97 Collab.), Nucl. Phys. A {\bf 661}, 130c (1999).
\bibitem{StatMod}
    P. Braun-Munzinger, J. Cleymans, H. Oeschler and K. Redlich,
    Nucl. Phys. A {\bf 697}, 902 (2002).
\bibitem{ramello}
	L.~Ramello, (NA50 Collab.), talk given at QM2002 
\bibitem{Spieles:1999kp}
C.~Spieles, R.~Vogt, L.~Gerland, S.~A.~Bass, M.~Bleicher, H.~St\"ocker and
W.~Greiner,
Phys.\ Rev.\ C {\bf 60} (1999) 054901
\bibitem{Cassing:1997kw}
W.~Cassing and E.~L.~Bratkovskaya,
Nucl.\ Phys.\ A {\bf 623} (1997) 570
\end{thebibliography}
\end{document}